%% file: main.tex
\documentclass[a4paper]{jpconf}
\usepackage{graphicx}
\input{preamble}

\begin{document}
\title{Evaluation of Portable Programming Models to Accelerate LArTPC Detector Simulations}


\author{Zhihua Dong$^1$, Kyle Knoepfel$^2$, Meifeng Lin$^{1}$, Brett Viren$^3$ and Haiwang Yu$^3$}

\address{$^1$Computational Science Initiative, Brookhaven National Laboratory, Upton, NY 11973, USA}
\address{$^2$Scientific Computing Division, Fermi National Accelerator Laboratory, Batavia, IL 60510, USA}
\address{$^3$Department of Physics, Brookhaven National Laboratory, Upton, NY 11973, USA}

\ead{zdong@bnl.gov, knoepfel@fnal.gov, mlin@bnl.gov, bv@bnl.gov, hyu@bnl.gov}

\begin{abstract}
The Liquid Argon Time Projection Chamber (LArTPC) technology is widely used in high energy physics experiments, including the upcoming Deep Underground Neutrino Experiment (DUNE). Accurately simulating LArTPC detector responses is essential for analysis algorithm development and physics model interpretations. Accurate LArTPC detector response simulations are computationally demanding, and can become a bottleneck in the analysis workflow. Compute devices such as General-Purpose Graphics Processing Units (GPGPUs) have the potential to substantially accelerate simulations compared to traditional CPU-only processing.  The software development for these compute accelerators often carries the cost of specialized code refactorization and porting to match the target hardware architecture. With the rapid evolution and increased diversity of the computer architecture landscape, it is highly desirable to have a portable solution that also maintains reasonable performance. We report our ongoing effort in evaluating Kokkos as a basis for this portable programming model using LArTPC simulations in the context of the Wire-Cell Toolkit, a C++ library for LArTPC simulations, data analysis, reconstruction and visualization.
\end{abstract}

\input{intro}

\input{tech}

\input{results}

\input{summary}

\input{refs}
\end{document}

%% file: preamble.tex
\usepackage[dvipsnames]{xcolor}
\usepackage[hidelinks]{hyperref}
\usepackage{graphicx}
\usepackage{wrapfig}
\usepackage{amsmath}

%% file: intro.tex
\section{Introduction}

\input{intro-portability}

\input{intro-simulation}

%% file: intro-portability.tex

The experimental particle physics community developing Liquid Argon Time Projection Chamber (LArTPC)
detectors today face the need to provide software
for simulation and data processing which can efficiently run on a
variety of hardware platforms.  The same software elements may run on a laptop,
a local GPU workstation, an institutional cluster, ``grid'' computing resources and
high performance computing (HPC) facilities.  The architectures for these systems are diverse, with CPUs and different flavors of GPUs, each of which may have a different native application programming interface that is favored by the hardware vendor. Such examples include  CUDA~\cite{cuda} for NVIDIA GPUs, HIP~\cite{hip} for AMD GPUs and SYCL~\cite{sycl} for Intel GPUs. Redeveloping unique software solutions for any given
algorithm so that they may run efficiently over this variety is labor
intensive.  We thus seek methods  to produce a single algorithm implementation that can be run on different hardware architectures.

%% file: intro-simulation.tex

As for all modern particle physics experiments, detector simulation
is critical for many facets of experiments utilizing LArTPC
technology.  Such simulation is needed to influence initial detector design decisions and final design
validation.  It also provides initial fodder for developing reconstruction
and analysis codes which must also perform correctly on real detector
data.  Likewise, researchers require methods to propagate systematic uncertainties
through simulation codes and relate them to results from real detector data.  In recent years,
simulation takes on the new, important and challenging duty of
supplying training data to high-precision artificial-intelligence
machine-learning procedures.  Each of these duties, particularly the
last, poses stringent requirements on the simulation for correctness
and on the computational resources needed to produce sufficient sample
sizes. It is thus very important to have an efficient, and preferably portable, LArTPC simulation implementation that can run on a variety of hardware platforms, including those equipped with GPUs. We thus investigate the best strategies to implement LArTPC simulations in a performant and portable manner across different hardware architectures, which we will further describe in the following sections.

%% file: tech.tex
\section{Technical Details}

\input{tech-signal}

\input{tech-wctgen}

\input{tech-kokkos}

%% file: tech-signal.tex
\subsection{LArTPC Signal Simulation}

\begin{wrapfigure}{r}{0.5\textwidth}
  \centering
  \includegraphics[width=0.5\textwidth]{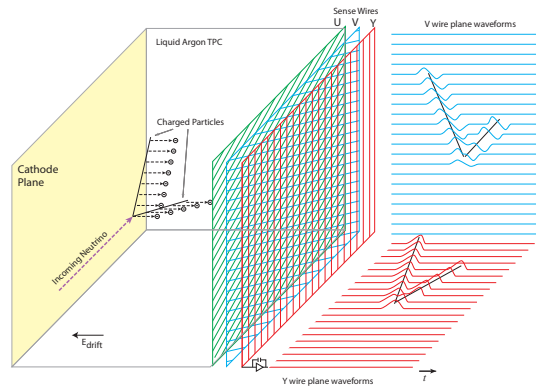}
  \caption{Illustration from Ref.~\cite{lartpc-sigform} of a three-wire plane LArTPC and its signal formation.}
  \label{fig:drift}
\end{wrapfigure}

Figure~\ref{fig:drift} illustrates several key features of the LArTPC detector
technology relevant to the processing we describe.  

Energetic
particles passing through the liquid argon will ionize electrons.
These electrons will drift through the applied electric field toward a
series of anode planes, each composed of a series of parallel wires
and each oriented at some angle with respect to the others.  The wire
planes are given an electrical potential which biases them relative to
this drift field so that electrons will pass by the wires of the first
planes and collect on wires of the last plane.  This results in
bipolar signals collected by readout electronics attached to the wires
of all but the last plane while that last plane produces unipolar
signals.

To greatly reduce computational complexity, the simulation treats each
plane independently which allows for ignoring the dimension along the
wire span.  The two-dimensional (2D) space of the problem then spans
the longitudinal or electron-drift direction and the transverse or
wire-pitch direction.  The simulation models signal formation as:

\begin{equation}\label{eq:conv}
  M(t,x) = {\int_{-\infty}^{\infty}\int_{-\infty}^{\infty} R(t-t',x-x') \cdot S(t',x')dt' dx'} + N(t,x), 
\end{equation}
Where $t$ is (discrete) sampling time, $x$ indicates sense wire
position along the transverse direction, $M(t,x)$ is the ``measured''
value such as what is input to an analog-to-digital converter (ADC)
channel $x$ at time $t$.
The term $R(t-t',x-x')$ represents the detector response to a unit
pulse of drifting ionization charge.
It is composed of terms covering the current induced in the sense
wires by that drifting and the response to that current by the shaping
electronics resulting in a voltage waveform.
The term $S(t',x')$ represents the charge distribution of the
ionization electrons over the longitudinal (time) and transverse (wire
pitch) dimensions.
Finally $N(t,x)$ represents voltage level output of various
electronics and coherent noise models.
%

%% file: tech-wctgen.tex
\subsection{Original LArTPC Simulation implemented in Wire-Cell Toolkit}
\label{sec:impl}

Wire-Cell Toolkit (WCT)~\cite{wct} is a C++ software package for LArTPC simulation, reconstruction and visualization.
In addition to algorithms, WCT features a dataflow programming (DFP) architecture as well as a modular design.
The current state-of-the-art LArTPC simulation has an implementation in WCT which is used by many experiments.
For the portable parallelization porting, we focus on the signal simulation part, which is the first term
in Eq.~\ref{eq:conv}, as it is more computationally demanding than the noise term $N(t,x)$.
In WCT, this 2D convolution is calculated
by applying the (fast) Fourier transform (FFT) on $S(t,x)$ to produce a 2D
array in the frequency (time/longitudinal and space/transverse)
domain, $\mathcal{S}(\omega_{t},\omega_{x})$.  Then, the similarly transformed 2D detector response array $R(\omega_{t},\omega_{x})$
is multiplied by $\mathcal{S}(\omega_{t},\omega_{x})$, and the convolution is completed by applying the inverse
Fourier transform back to the time-space domain.

\begin{align}\label{eq:conv-freq}
\begin{split}
  S(t,x) \quad &\underrightarrow{\quad {\rm FT} \quad} \quad \mathcal{S}(\omega_{t},\omega_{x}), \\
  M(\omega_{t},\omega_{x}) &= R(\omega_{t},\omega_{x}) \cdot S(\omega_{t},\omega_{x}), \\
  M(\omega_{t},\omega_{x}) \quad &\underrightarrow{\quad {\rm FT^{-1}} \quad} \quad M(t,x),
\end{split}
\end{align}

The inputs to WCT LArTPC simulations are \textit{ionization electron groups} sampled at fixed space and time coordinates from GEANT4 simulations~\cite{geant4}. After simulating the diffusion process along the drifting, each ionization electron group is diffused into a 2D distribution along the drifting and wire pitch directions. The following simulations consist of three key computational tasks and will be the focus of this paper:
\begin{itemize}
    \item \textbf{Raster}: Individual distributions of ionization electron groups are binned, each as a ``patch'' of varying sizes and typically of $\mathcal{O}(20\times20)$ elements.
    \item \textbf{Scatter-add}: These patches are stacked and summed over a larger grid which spans the longitudinal time and transverse space of the union of the patches, an array of $\mathcal{O}(10,000 \times 10,000)$ elements.
    \item \textbf{Convolution}: The large grid is convolved with the detector responses as described above.
\end{itemize}

These three tasks all have non-negligible contributions to the total time consumption in the LArTPC signal simulation. The exact proportion of time consumption for each task depends on the simulation parameters. We will see in Section~\ref{sec:perf}, in the data set we tested, \textit{Rasterization}, \textit{Scatter-Add} and \textit{FFT} each consume about 67\%, 5\% and 22\%, respectively, of the total kernel run time with the original unimproved CPU code. 

%% file: tech-kokkos.tex
\subsection{Kokkos-based version}
Kokkos~\cite{kokkos} is a C++ library that provides an abstraction layer to achieve 
performance portability. It supports several different node
architectures and memory models by allowing users to define their own
execution and memory spaces.  It maps C++ source code to different
\textit{backends} during build time to achieve portability across different architectures. Kokkos currently supports OpenMP or POSIX threads (pthreads) for multithreading on CPUs, CUDA for NVIDIA GPUS, HIP for AMD GPUs, OpenMP target
offloading and SYCL, the latter two of which are also portable programming models themselves. Due to its extensive support of different architectures, we chose Kokkos as the first portable programming model to investigate. 

\begin{figure}
  \centering
  \includegraphics[width=.9\columnwidth]{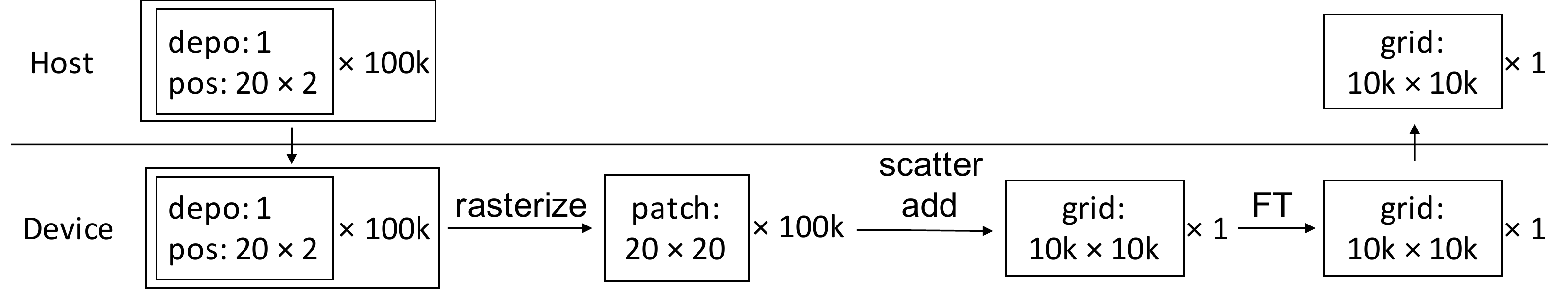}
  \caption{Data flow for the full Kokkos porting. Numbers following the colons (``:'') indicate the number of elements in that object before the colons.}
  \label{fig:data-flow}
\end{figure}

To better experiment with the Kokkos abstraction layer, we developed the standalone Wire-Cell-Gen-Kokkos module~\cite{wct-gen-kokkos}. In our vCHEP21 presentation and proceedings~\cite{vchep21}, we showed results for a partial porting as a demonstration. Here we report the results of a full porting that implements all the main computational tasks in Kokkos. Figure~\ref{fig:data-flow} shows the data flow design between host (CPU) and device (GPU) of the full porting. This design minimizes the data transfer needed between memory spaces. To achieve this  data flow design, we did a major overhaul of the data objects with an overall direction from arrays of structures/objects to structures/objects of arrays. Please note, this data object refactoring is not a simple one-to-one mapping, so that the interfaces to algorithms are changed, which fit better to parallelization. The new data objects enabled parallelization at the level of individual ionization electron groups which exposes much higher concurrency compared to patch sample point level (100k vs. 400). More details on this can be found in~\cite{vchep21}. In addition to changes in data objects, we list several other key implementation details below. Please refer to Ref.~\cite{wct-gen-kokkos} for more details.

\begin{itemize}
    \item \textbf{Extensions to the toolkit:} We added a C++ \texttt{KokkosEnv} context manager
  component to initialize and finalize Kokkos as well as special build
  system support.

\item \textbf{Using dense matrix in ``Scatter Add'':} Compared to original sparse representation, using dense matrix increases speed even in single threaded execution.

\item \textbf{FFT Wrapper API:} Since Kokkos does not provide an API to optimized vendor FFT libraries (\texttt{FFTW}, \texttt{cuFFT},  \texttt{rocFFT}, etc.), we implemented our own FFT wrapper similar to that of the Synergia group~\cite{synergia}. 
\end{itemize} 

%% file: results.tex
\section{Results}

\input{results-validation}

\input{results-architecture}

\input{results-multiproc}

%% file: results-validation.tex
\subsection{Correctness Validation}

\begin{figure}
  \centering
  \includegraphics[trim=4cm 1cm 3cm 2cm, clip=true, width=.8\columnwidth, height=2.5in]{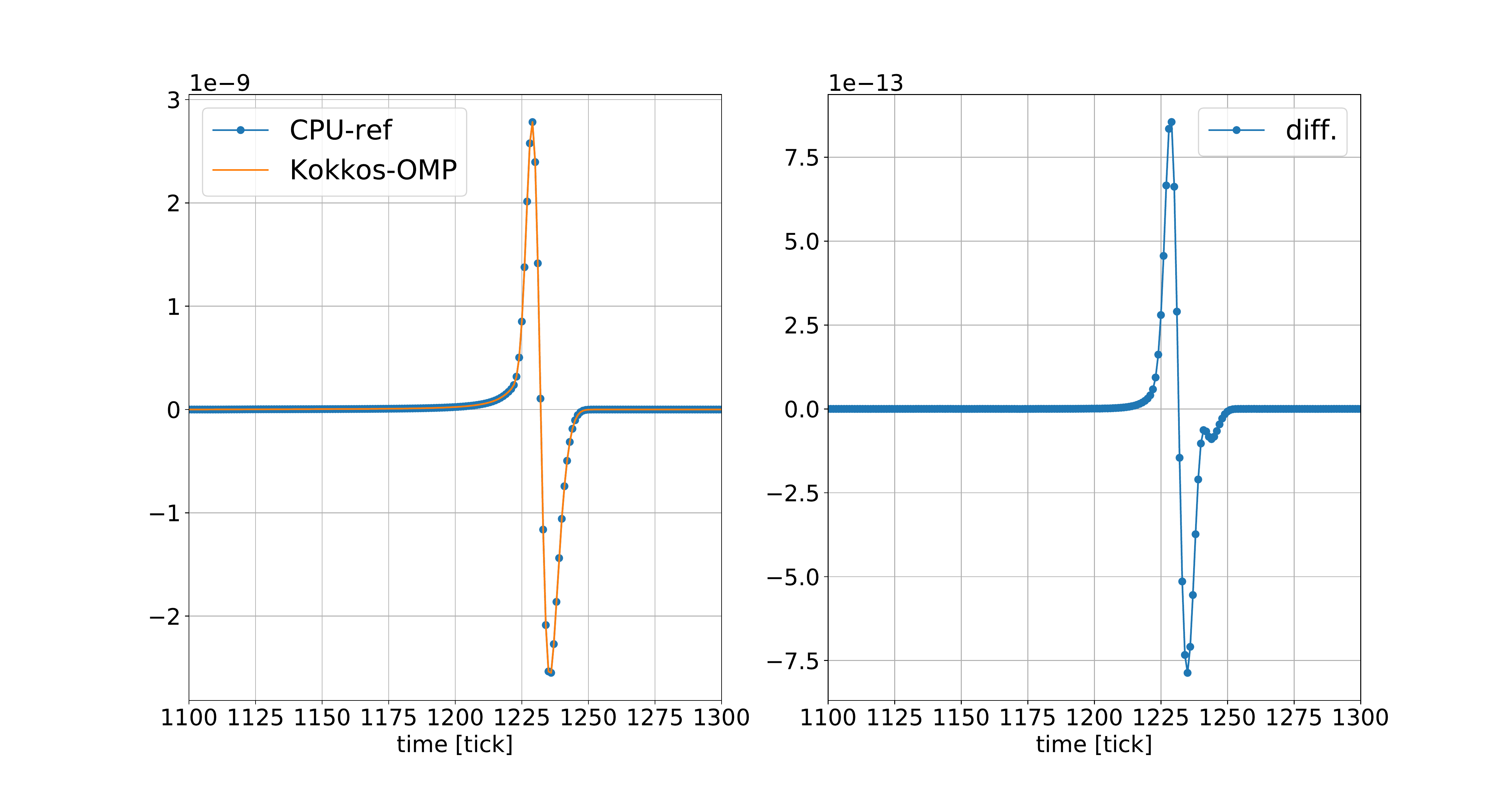}
  \includegraphics[trim=4cm 1cm 3cm 2cm, clip=true, width=.8\columnwidth, height=2.5in]{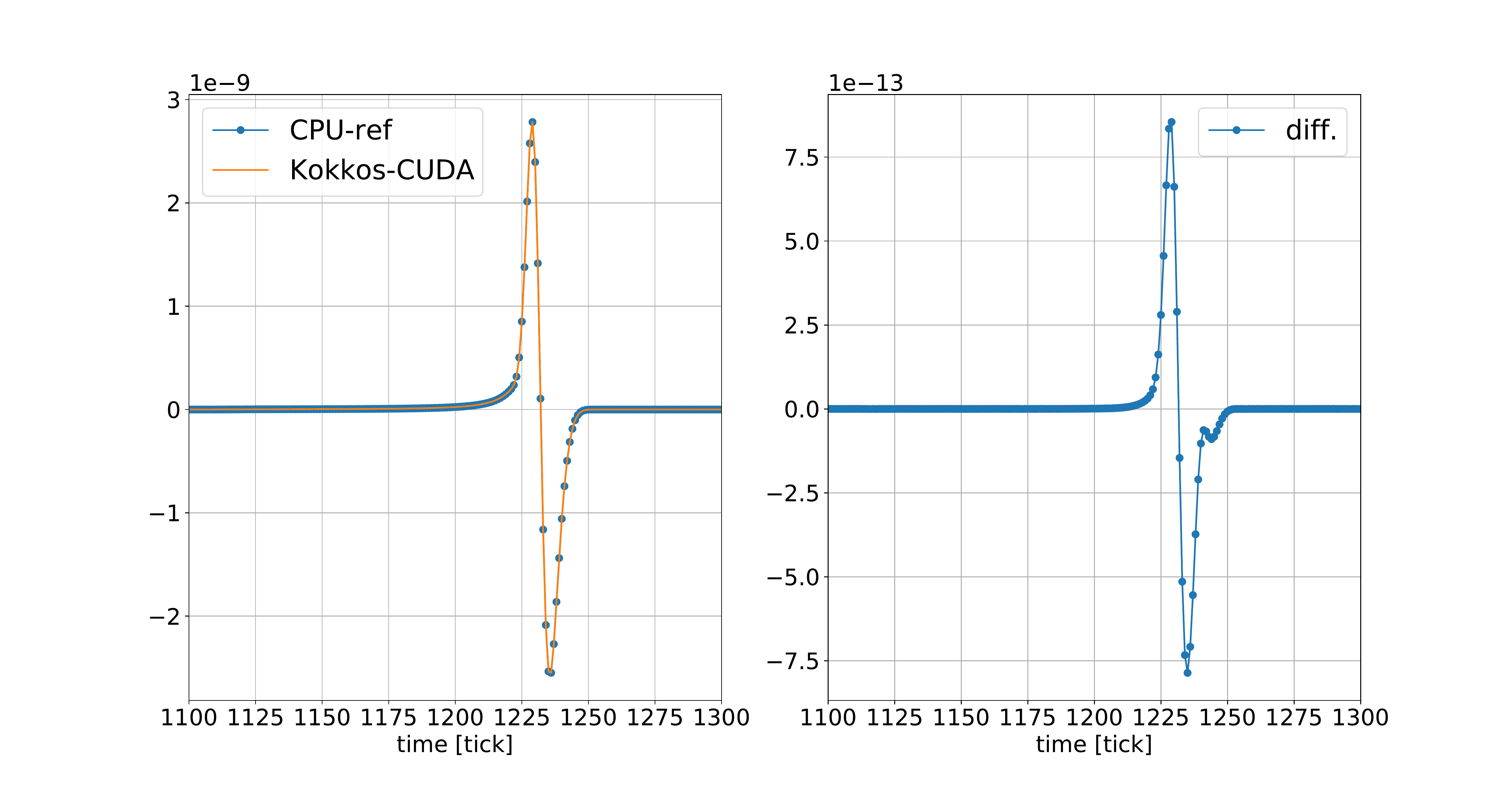}
  \caption{Waveform comparisons from CPU-ref and Kokkos with OpenMP (Top) and CUDA (Bottom) backends. Y-axes are related to induced currents before application of the electronic response. The units are the same for all plots.}
  \label{fig:wave-comp}
\end{figure}

Simulated waveforms from the original CPU reference implementation (CPU-ref) are compared to those produced by the Kokkos implementation with different backends in Figure~\ref{fig:wave-comp}.
For both CUDA and OMP backends, the differences are at the 0.01\% level,
which are likely caused by different samplings of the random number streams, as the original CPU version uses binomial distribution while normal distribution approximation is used in the current Kokkos implementation. 
As such the results are not expected to agree exactly, and the differences we see are not statistically significant. However, we do observe some kinks in the \texttt{diff} plots, which we still need to investigate further. 

%% file: results-architecture.tex
\subsection{Performance on Different Architectures.}
\label{sec:perf}
To check how the CUDA, HIP and OMP backends of our Kokkos implementation perform on respective architectures, we  obtained timing information, averaged over 10 runs, for the three key computational tasks as discussed in Section~\ref{sec:impl}. The CUDA and OMP backends were tested on the NERSC Perlmutter system~\cite{perlmutter}, each node of which has 4 NVIDIA A100 GPUs  and an AMD EPYC 7763 64-core CPU. For this test, only one A100 GPU was used.  The HIP backend was tested on a local workstation (Lambda1) at Brookhaven National Laboratory which has an AMD Raedon Pro VII GPU and an AMD 24-core Ryzen Threadripper 3960X CPU. On both systems, we also ran the single-threaded reference CPU version (CPU-ref) as comparison. The detailed timing information is shown in Table~\ref{tab:timing}.  On average, running on one A100 GPU with the Kokkos-CUDA backend is about 33 times faster than CPU-ref, while the Kokkos-HIP version runs about 9 times faster than CPU-ref. However, the OpenMP backend with 64 CPU threads (Kokkos-OMP64) is only marginally faster than the original CPU code. This is due to the fact that the CPU version of the FFT is not parallelized. The slowdown from Kokkos-OMP64 is expected as it performs larger FFTs than CPU-ref. We are still working to optimize this part. 
\vspace{4mm}
\begin{table}[htbp]
    \centering
    \begin{tabular}{|l|c|c|c|c|c|}
    \hline
    System & \multicolumn{3}{c|}{Perlmutter} & \multicolumn{2}{c|}{Lambda1} \\ 
    \hline 
    Backend & CPU-ref& Kokkos-CUDA & Kokkos-OMP64 & CPU-ref & Kokkos-HIP \\ 
        \hline
        \hline
Rasterization [secs]  &   12.3   & 0.048  & 0.08  & 10.5 & 0.072\\
ScatterAdd [secs] & 1.04 &     0.00045  & 0.014 &  1.05 &  0.0066 \\
FFTs  [secs]    & 3.95 &       0.30  & 9.71    &  5.01 & 1.49 \\
\hline
Total [secs] &  18.31 &   0.55 &  10.55 & 17.55  & 1.87 \\
\hline
    \end{tabular}
    \caption{Main computational task times on tested architectures (10-run average) with the Kokkos implementation and the reference CPU implementation.}
    \label{tab:timing}
    \end{table}

%% file: results-multiproc.tex
\subsection{Performance with Multiple Processes.}

\begin{wrapfigure}{r}{0.65\textwidth}
    \centering
    \includegraphics[width=0.65\textwidth]{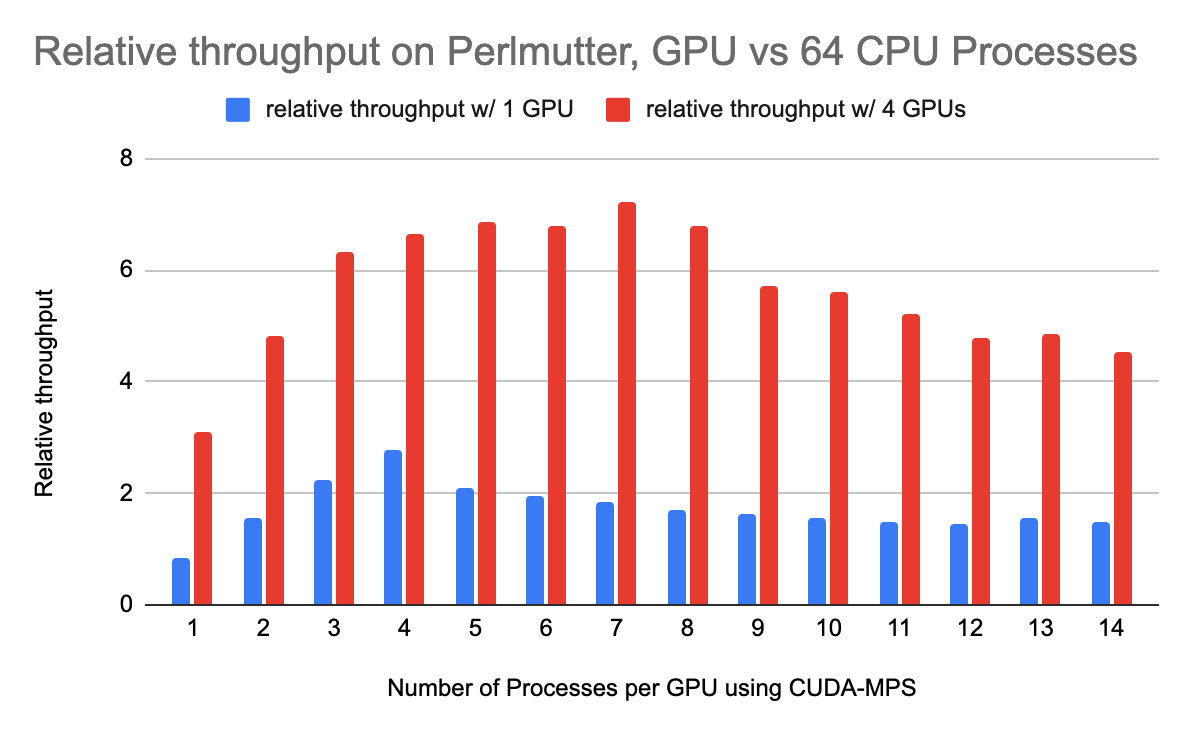}
    \caption{Throughput of Kokkos-CUDA implementation on Perlmutter relative to that of the reference implementation using all 64 CPU cores.}
    \label{fig:throughput}
\end{wrapfigure}
The timing results in Table~\ref{tab:timing} are from running one single process. We can also run multiple independent processes to share the GPUs and increase total throughput on the GPUs. This is analogous to the traditional high-throughput computing model, where a multi-core CPU is shared by multiple independent processes. Figure~\ref{fig:throughput} shows the relative throughput achieved on Perlmutter with the Kokkos-CUDA backend to share one or all four GPUs on the node, compared to running 64 CPU processes with the CPU-ref code. We see that the relative throughput increases as we increase the number of processes per GPU until the peak is reached, indicating that with one process per GPU the GPU is under-utilized. As the number of processes increases further, the overall throughput actually degrades. This is likely due to CPU clock throttling as more processes, and therefore more CPU cores, are running. We have observed this throttling behavior in our CPU-only test (not shown in this paper).  We can achieve up to 7 times more throughput with 4 GPUs and 3 times more with 1 GPU compared to a fully loaded CPU. Why the throughput with 4 GPUs is not four times more than that with 1 GPU is still under investigation. 

%% file: summary.tex
\section{Summary and Outlook}


We have implemented the LArTPC signal simulation in Kokkos as a portable parallelization solution. A major refactoring was done to increase concurrencies for better parallelization performance. Current Kokkos implementation with the CUDA backend achieved significant single GPU speedup and node level throughput increase compared to the original single-thread CPU version. Going forward we intend to investigate further optimizations as well as other portable programming models, such as OpenMP and SYCL.
 

\ack{This work was supported by the U.S. Department of Energy, Office of
Science, Office of High Energy Physics, High Energy Physics Center for
Computational Excellence (HEP-CCE) under B\&R KA2401045.  This research used resources of the National Energy Research Scientific Computing Center (NERSC), a U.S. Department of Energy Office of Science User Facility located at Lawrence Berkeley National Laboratory, operated under Contract No. DE-AC02-05CH11231. We gratefully acknowledge the support of the Wire-Cell team of the Electronic Detector Group in the Brookhaven National Laboratory Physics department, both of which are supported by the U.S. Department of Energy under Contract No. DE-SC0012704.}